\documentclass[pdflatex,sn-mathphys-num]{sn-jnl}

\usepackage{graphicx}%
\usepackage{multirow}%
\usepackage{amsmath,amssymb,amsfonts}%
\usepackage{amsthm}%
\usepackage{mathrsfs}%
\usepackage[title]{appendix}%
\usepackage{xcolor}%
\usepackage{textcomp}%
\usepackage{manyfoot}%
\usepackage{booktabs}%
\usepackage{algorithm}%
\usepackage{algorithmicx}%
\usepackage{algpseudocode}%
\usepackage{listings}%

\usepackage[pagewise]{lineno}

\theoremstyle{thmstyleone}%
\theoremstyle{thmstyletwo}%

\theoremstyle{thmstylethree}%

\newcommand{\Msun}{M$_\odot$}
\newcommand{\ion}[2]{\mbox{#1\,\textsc{#2}}}

\begin{document}

\title[Article Title]{Tracing the s-Process: Spectroscopic Insights into Chemical Abundances in O- and C-rich Evolved Stars}


\author*[1,2]{\fnm{S.} \sur{Van Eck}}
\email{sophie.van.eck@ulb.be}
\equalcont{These authors contributed equally to this work.}

\author[3]{\fnm{C.} \sur{Abia}}
\email{cabia@ugr.es}
\equalcont{These authors contributed equally to this work.}

\author[3]{\fnm{I.} \sur{Dom\'\i nguez}}
\equalcont{These authors contributed equally to this work.}

\affil[1]{\orgdiv{Institut d’Astronomie et d’Astrophysique},\orgname{Université libre de Bruxelles}, 
\orgaddress{\street{CP 226}, \postcode{1050} \city{Brussels},  \country{Belgium}}}

\affil[2]{\orgdiv{BLU-ULB}, \orgname{Brussels Laboratory of the Universe, \url{https://blu.ulb.be}}}

\affil[3]{\orgdiv{Dpt. F\'\i sica Te\'orica y del Cosmos}, \orgname{Universidad de Granada}, \orgaddress{\street{Avda. Fuentenueva s/n}, \city{Granada}, \postcode{18071}, \country{Spain}}}


\abstract{Oxygen and carbon-rich AGB stars - and objects directly polluted by them - are excellent laboratories to investigate the nucleosynthesis and mixing processes occurring during the later phases of the of low- and intermediate-mass star evolution. The determination of the abundances of several s-elements is a key tool for constraining theoretical AGB models. This contribution discusses the main results, recent advances, and current problems on this subject.}

\keywords{AGB stars, stellar abundances, nucleosynthesis, s-process}

\maketitle

\section{Introduction}\label{sec1}

Evolved giant stars, or "red stars" as they were called before the XX$^{\rm th}$ century, have long been recognised as an important potential site for the production of chemical elements in the Universe. Some of these objects are now understood to belong to the red giant branch (RGB) or asymptotic giant branch (AGB) phases of evolution, based on their location in the Hertzsprung-Russell (HR) diagram and the comparison of observables with stellar evolution and nucleosynthesis models. However, they were once only stars of the M, S, SC, R, N, or C types.
The Italian astronomer Angelo Secchi, shortly after the application of spectroscopy to starlight, observed stars that he classified as "type IV", "faint" and showing "bloody red" colours. He judiciously noted that their spectra showed "a marked analogy with the reverse spectrum of carbon" and published wonderful drawn spectra of these \cite{Secchi-1868b} quoted in \cite{Houziaux-1979, McCarthy-1994}. 
The strong resonance lines of strontium and barium (\ion{Sr}{ ii} $\lambda$ 4077, 4215 \AA; \ion{Ba}{ii} $\lambda$ 4554 \AA) were then detected in the stars at the end of the XIX$^{\rm th}$ century (already detected by Secchi, but then also by Kirchhoff \& Bunsen, Lockyer, and Huggins), and 
during the first quarter of the XX$^{\rm th}$ century
\cite{Merrill-1922}, abnormally strong barium lines in Barium and S-type stars were recognised \citep{hea14}.
How these elements were produced was still debated until Merrill detected technetium lines in the spectra of S stars \cite{Merrill-1952}. This element has no stable isotope, and $^{99}$Tc, the one on the s-process path, has a half-life of only 2.1$\times 10^5$ years, i.e., short with respect to 
main-sequence lifetimes, proving that AGB stars were capable of nucleosynthesising their own heavy elements.

The AGB phase marks the stage of the most dramatic surface chemical changes of stars with an initial mass ranging from 0.8 to 8-10 M$_\odot$, driven by thermal pulses (TP), which are He-shell thermal instabilities that trigger deep convective mixing events known as third dredge-up (TDU). These events transport He-burning products, especially carbon, to the surface, mainly increasing the carbon content in the envelope, eventually turning the star from an O-rich star\footnote {The first dredge-up already affects the C/O in the opposite direction, decreasing it from its initial solar ratio (C/O)$_\odot\sim 0.5$, down to roughly 0.4 (M$\sim 1$~\Msun) or 0.3 (M$\ge $2~\Msun), actually due to an increase of the [N/C] ratio \cite[e.g.][]{Masseron-2015, kar14}.} (C/O$ <1$, abundance ratio by number) to a C-rich (C/O $\ge 1$), giving rise to a so-called carbon star \cite{Straniero-2023}. Heavy elements are synthesised during the same phase and dredged-up to the stellar surface, where they are detectable spectroscopically. 
They are then ejected into the interstellar medium through mass-loss processes driven by stellar pulsation, shock waves, and radiation pressure on dust grains. These enriched winds contribute to the chemical enrichment of the Galaxy.
The amount and distribution (among the atomic species) of the enrichment are sensitive probes of nucleosynthesis conditions and mixing processes within AGB stars. 
Extensive observations and measurements on several sources at the end of the last century, from low-metallicity stellar objects to presolar grains, from normal AGB stars to post-AGB sources and binary systems, led to the crucial paradigm change from the  $^{22}$Ne$(\alpha,$n$)^{25}$Mg to the $^{13}$C$(\alpha,$n$)^{16}$O nuclear reaction as the main neutron source 
 for the synthesis of the s-elements in AGB stars
 \cite{Busso-1999, Goriely-2018}. Unfortunately, the activation of such a neutron source is not straightforward. Current theoretical AGB models show that little $^{13}$C is left behind by shell H-burning in the zone surrounded by the two alternatively burning AGB star shells,
 as it is efficiently converted into 
$^{14}$N via CN cycling. To produce the necessary $^{13}$C, models invoke a partial mixing of protons from the convective envelope into the $^{12}$C-rich intershell during TDU episodes, leading to the formation of the so-called '$^{13}$C pocket'. Within this region, $^{13}$C is produced and later serves as the main neutron source through the $^{13}$C$(\alpha,$n$)^{16}$O reaction.  
Despite the uncertainties inherent to this mechanism,
modelling of the s-process, as provided by this main neutron-producing reaction, has been crucial in clarifying the origin and distribution of nuclei from Sr up to Pb and Bi in the Galaxy  \cite{Kappeler-2011,kar14,Cristallo-2015}. 
Subsequent stellar winds eject this processed material into the interstellar medium (ISM), enriching future generations of stars and their potential planets. Thus, AGB stars are one of the main engines of the chemical evolution of the universe \cite{pra20}.

For more than 25 years, a growing number of abundance determinations performed from high-resolution spectra and using sophisticated analysis techniques, included when applied to large survey data, have led to several refinements of the theory of heavy element nucleosynthesis, as first exposed in \cite{bur57}. The late Roberto Gallino (Fig. 1) played an important role in this field by encouraging stellar spectroscopists to search for insights into the s-process in the atmospheres of cool red giants. In this review paper, a short scientific tribute to his memory, we present the main recent advances coming from oxygen-and carbon-rich AGBs, or objects directly polluted by them (called extrinsic stars). In Sect 2 we discuss the observational world of the O-rich AGB stars and their s-process abundances, whereas in Section 3, we describe the observational status of AGB carbon stars. A summary and expected future achievements are detailed in Section 4.

\begin{figure}[ht]
\centering
\includegraphics[angle=-90,width=1.0\textwidth]{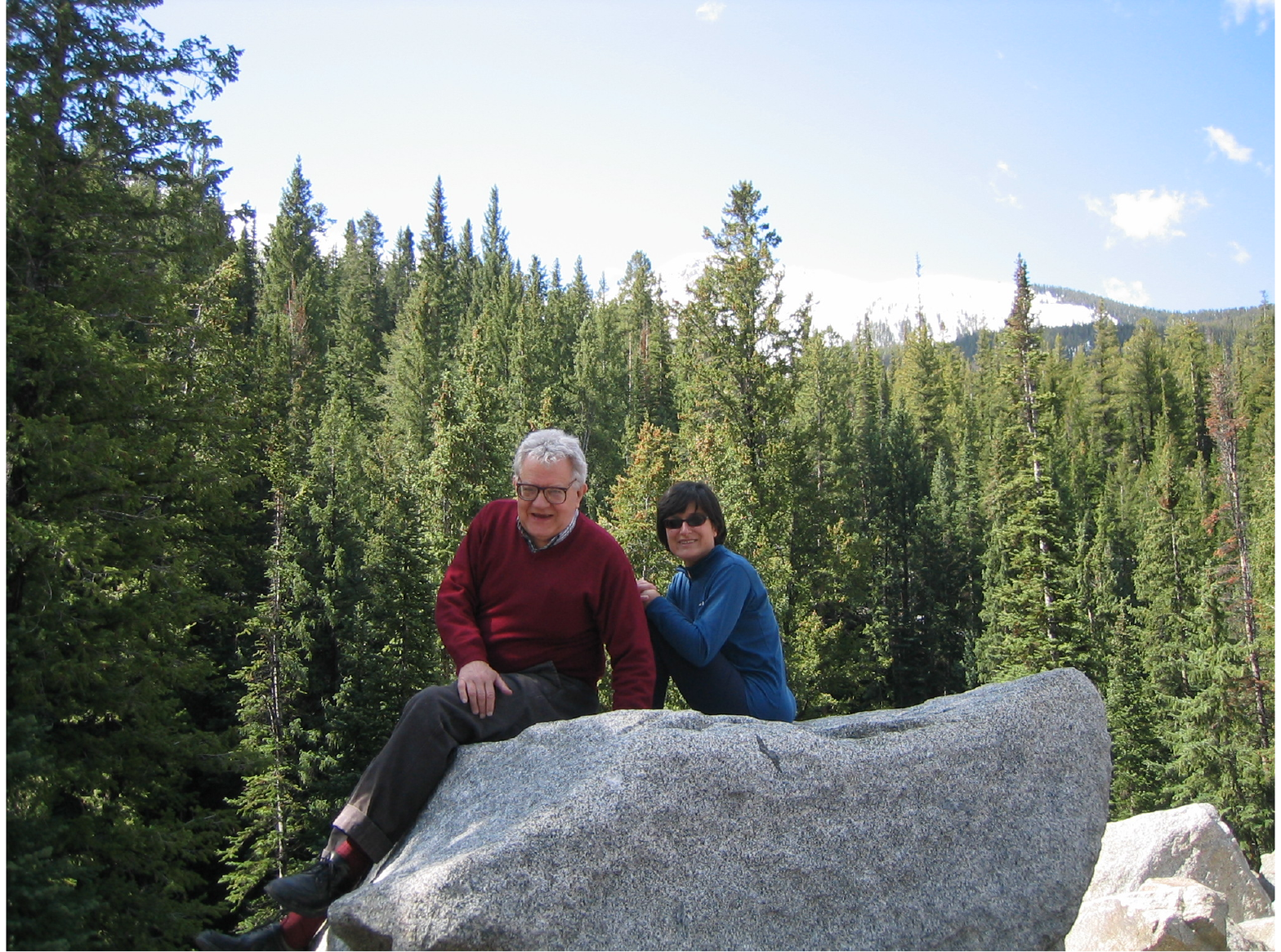}
\caption{Roberto Gallino and Inma Dom\'\i nguez during a weekend excursion at The Independence Pass during the summer workshop on the {\it Physics of the s-process} at the Aspen Center of Physics (2005).}\label{fig1}
\end{figure}

\section{AGB oxygen-rich stars}\label{sec2}

\subsection{The M-S-C sequence}
The M giants are cool enough to show prominent TiO  (for T$_{\rm eff} \lesssim 4000$K) and VO (for T$_{\rm eff} \lesssim 3000$K) bands in their optical spectra \cite{Keenan-1980,Gray-2009}. Their spectra and correspondence with stellar parameters have been studied in \cite{Fluks-1994, Bessel-1998}, and they have temperatures in the range of 2700-4000 K.

The many dredge-ups that occur on the TP-AGB periodically add s-process and C-rich matter to the stellar surface. 
Because CO has an exceptionally high dissociation energy ($\sim 11$ eV), molecular equilibrium in late-type stellar atmospheres strongly favours CO formation; hence, CO forms first and captures essentially all limiting elements (C or O), leaving the atmospheric chemistry of M giants (resp., C giants) dominated by the remaining excess oxygen (resp., carbon).

\begin{figure}[ht]
\centering
\includegraphics[width=\textwidth, trim=3cm 9cm 3cm 9cm, clip]{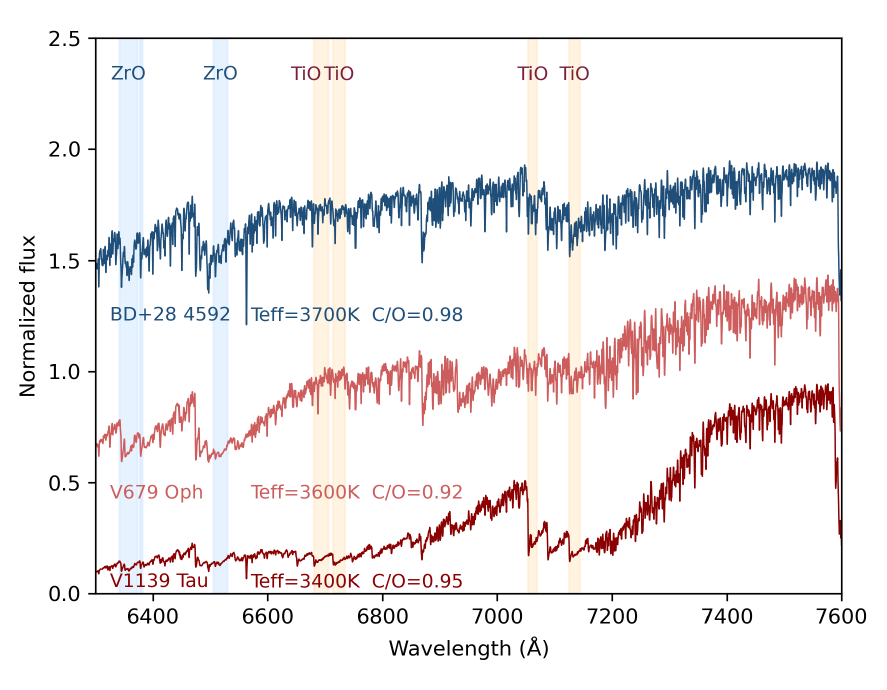}
\caption{Example HERMES \cite{Raskin-2011} spectra of three S-type stars analysed in \cite{Shetye-2021}, ordered by decreasing effective temperature as indicated. The spectra are normalized in the vicinity of 7550~\AA\ and vertically offset by 0.5 units for clarity. Selected TiO and ZrO molecular bands are marked.}\label{Fig:figS1}
\end{figure}

Due to the enrichment of the s-process elements (and not only because of the cooler surface temperature, as already demonstrated in \cite{Piccirillo-1977}), the ZrO bands become apparent, and therefore the star is classified in the MS or S family (Fig~\ref{Fig:figS1}). Actually, two types of S- and C-stars, with quite different evolutionary histories, co-exist: the intrinsic and extrinsic ones.
Intrinsic and extrinsic late-type giants represent two distinct evolutionary pathways towards chemical peculiarity among S and C stars. Intrinsic S and C giants are currently undergoing TPs and TDUs on the TP-AGB; therefore, their chemical enrichment is a self-produced signature of internal nucleosynthesis and mixing. In contrast, extrinsic giants, such as extrinsic S and C stars, may show similar overabundances but are not luminous or cool enough to be classified as TP-AGB stars. The extrinsic red giants are discussed in Sect.~\ref{Sect:ext}.

From the samples analysed by \cite{Shetye-2018}, intrinsic (Tc-rich) S stars occupy the TP-AGB phase at luminosities of approximately $3000-10\,000$ L$_{\odot}$, whereas extrinsic (Tc-poor) S stars lie in the upper RGB or early-AGB phase with typical luminosities of $\sim 1000$ L$_{\odot}$, rarely exceeding 5000 L$_{\odot}$.
Extrinsic and intrinsic S star  
masses range roughly from 1 to 3.5 M$_{\odot}$,
as inferred from their positions in the HR diagram by comparison with stellar evolution models \cite{Shetye-2018, Shetye-2019}.
S stars have effective temperatures in the same range as those of M-type stars, roughly from 2700 K to 4000 K, but they are characterised by a C/O ratio mostly larger than the solar value\footnote{Solar C/O$=0.59$ (by number: C/O = $10^{\rm A(C)-A(O)}$) \cite{Asplund-2021}, however, this lower limit for the S star C/O ratio can be affected by the first dredge-up: C/O is predicted to drop to 0.44 (resp., 0.33) in a solar metallicity, 1 M$_{\odot}$ (resp., 2 M$_{\odot}$) star \cite{kar14}.}
but strictly smaller than 1. The S giants are often mentioned to have C/O$=1$, but this is an erroneous statement. 
Stars with C/O$=1$ are actually named SC stars, and they have a very peculiar undepressed optical spectrum, linked to the fact that all C and O atoms are trapped in the strongly bound CO molecule, leaving none for TiO, ZrO, VO, LaO and other metallic oxide molecular species \cite{Abia-1998}.
Finally, giants of the S spectral type should not be confused with objects that have been observed orbiting Sgr A* and unfortunately labelled as {\it S-$0n$ or S$n$ stars} (where $n$ is a running number)\cite{Reid-2009} nor, obviously, with {\it S-type} planets (orbiting one component of a binary system).

Several spectroscopic studies have confirmed the M-S-C sequence: these stars indeed share a common relation of increasing s-process enhancements with increasing $^{12}$C: C/O $\sim 0.46 \pm 0.13$ for M stars, \cite[see also \citealp{Smith-1990}]{Nandakumar-2024a}, 
$0.5 <$ C/O $< 1$ in S-type stars, $\sim 1$ in SC or CS stars \cite{Vaneck-17,Shetye-2018} and $>1$ in C stars \cite{abi02}, each group exhibiting a large scatter.

The mass loss rates are also compatible with the M-S-C sequence, since the S star mass loss rates and expansion velocities appear very similar to those of M stars and slightly lower than those of C stars ($2\times 10^{-7}$ M$_\odot$ yr$^{-1}$), while the highest rate (up to $\sim 10^{-4}$ M$_\odot$ yr$^{-1}$) characteristic of dust-enshrouded AGBs are found more frequently among C stars \cite{Ramstedt-2006,Guandalini-2010,Groenewegen-2018}.

\begin{figure}[ht]
\centering
\includegraphics[width=0.99\textwidth]{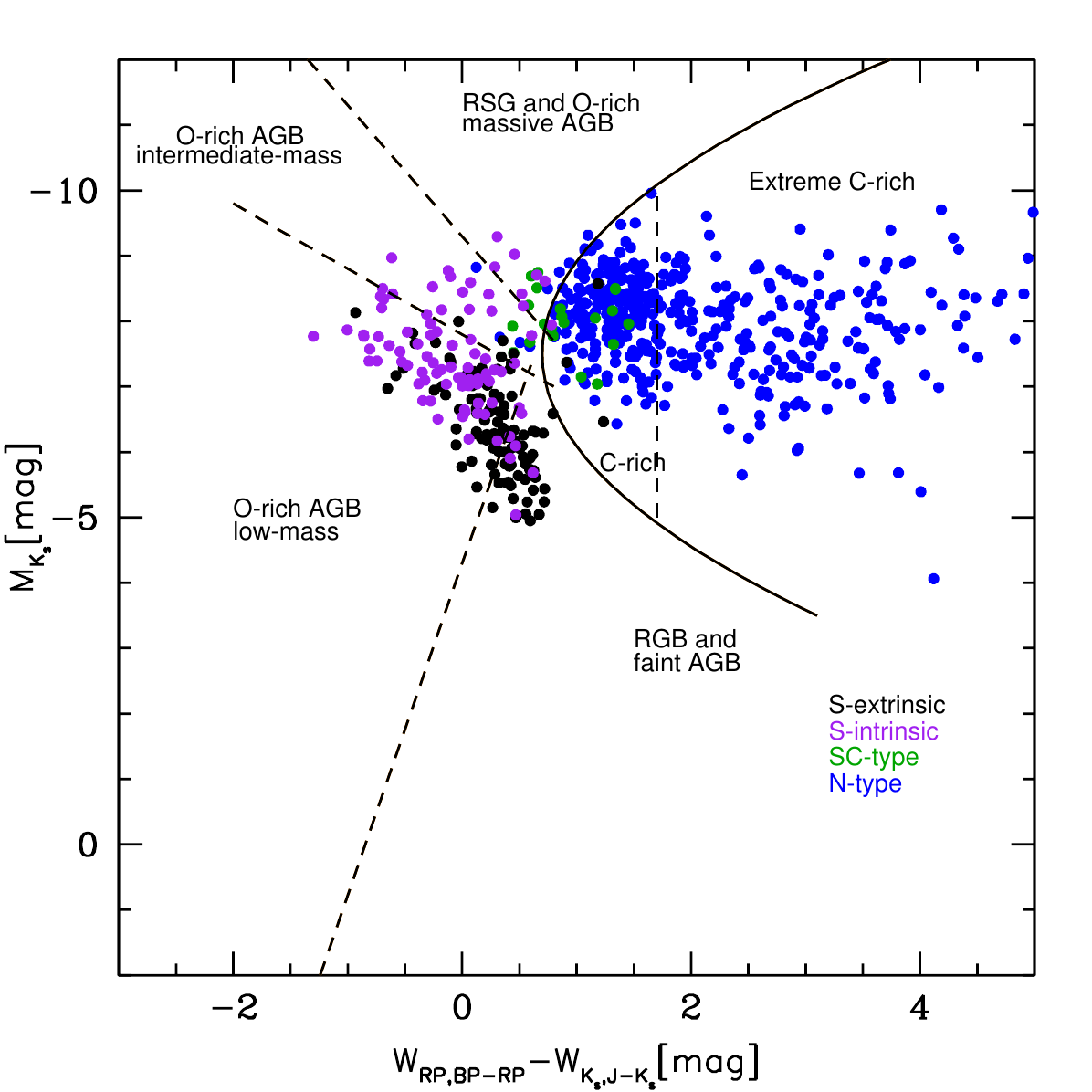}
\caption{ The Gaia-2MASS diagram for Galactic O-rich and C-rich AGB stars taken from the sample in \cite{abi22}. Clearly both types of stars occupy different areas in the diagram: intrinsic O-rich S stars (purple dots) are in average brighter than the extrinsic S stars (black dots). Many of SC-type stars (C/O$\approx 1$, green dots) occupy the transition region between carbon and oxygen rich objects, while N-type (carbon rich) stars (blue dots) are the reddest objects in the diagram (see \cite{leb18} for a detailed explanation of the different regions).}\label{fig3}
\end{figure}

The M-S-C sequence is difficult to prove from the location of the stars in the HR diagram because of the degeneracy between mass, metallicity, and evolutionary phase, and because of the presence of binary intruders. This degeneracy can also be seen in the  Gaia-2MASS diagram (see, e.g. Fig. 3 in \cite{leb18} and Fig.~\ref{fig3}). This diagram, originally introduced by \cite{leb18}, is particularly suitable for highlighting the presence of AGB stars. In this diagram, the M$_{K_{s}}$ magnitude is correlated with a particular combination of Gaia and 2MASS (Two Micron all Sky Survey) photometry through the quantity W$_{\rm{RP,BP-RP}}-$W$\rm{_{K_{s},J-K_s}}$, where W$_{\rm{RP,BP-RP}}$ and W$_{\rm{K_{s},J-K_s}}$ are almost reddening-free Wesenheit functions \citep{sos05}. This diagram is a powerful tool for analysing and identifying  different spectral types of AGB stars as a function of their chemical type and initial mass. The diagram also distinguishes between regions with low-, intermediate-mass, and massive O-rich stars \cite{leb18} (see Fig.~\ref{fig3}).

\subsection{Chemical abundances in O-rich AGB stars}
Although hydrodynamical model atmospheres have been developed for oxygen-rich AGB stars (e.g. \cite{Bladh-2019, fre23}), they have mostly been used to derive mass loss rates or photospheric properties. Abundance studies still heavily rely on 1-dimensional hydrostatic model atmospheres and therefore focus on the least variable objects. Pulsation and large-scale convection can lead to significant line-profile distortions and line doubling, as demonstrated by atmosphere tomography studies (e.g. \cite{Kravchenko-2018, Kravchenko-2019}), which makes any detailed spectroscopic analysis particularly challenging in the most variable objects.

\subsubsection{M stars}
Abundances in M-type stars have been investigated from the optical and infrared (e.g., \cite{Hayes-2022, Nandakumar-2024a, Nandakumar-2024b}). They show a roughly solar C/O ratio and no other heavy-element abundance trend\footnote{Except maybe for 
barium, since some moderate metal-poor M giants seem to have increased Ba abundances.} than the one originating from Galactic chemical evolution and also derived in FGK dwarf stars of the same metallicities from spectral lines at visual wavelengths \cite{DelgadoMena-2017,Tautvaisine-2021}.

\subsubsection{Tc-rich M stars}

M giants rich in Tc, covering a metallicity range $-0.9 \leq$ [Fe/H] $\leq -0.3$\footnote{We adopt the usual  notation  [X/Y]$\equiv$ log  (X/Y)$_*-$ log  (X/Y)$_\odot$ any element abundance ratio X/Y by number respect to the Sun. In the following, we refer with ls to the light mass $s$-elements Sr, Y and Zr and, with hs, to the high-mass $s$-elements Ba, La, Ce, and Nd. However, the Nd abundance in the solar system has a significant contribution from the r-process \cite{pra20}.} have been identified by \cite{Little-Marenin-1979, Little-Marenin-1987, Vanture-1991, Uttenthaler-2010, Shetye-2021, Shetye-2025}. Given the broadened lines of these stars, probably due to strong convective motions in their atmospheres, Tc abundances can hardly be derived, but Tc detection is unambiguous because the diagnostics from two \ion{Tc}{i} lines systematically agree. In these stars, the Tc lines become detectable before the ZrO bands show any significant enhancement; therefore, they are still classified as M-type stars. Their small C/O ratio ($0.3 < $C/O$ < 0.5$) indicates that they have experienced only one or a few TDUs, as confirmed by low-mass stellar evolution models (1.5 M$_{\odot}$) with subsolar metallicity ([Fe/H] = –0.5), with which they are marginally compatible. 
Abundance determinations are difficult in these highly variable stars. Tc-rich M stars exhibit stronger TiO bands than S stars, while displaying ZrO bands of comparable strength.
A possible slight enrichment in Zr, which would normally enhance the 
ZrO bands, 
may therefore remain unnoticed if it occurs  in the presence of very strong TiO bands, leading the star to be classified as an M star rather than as a S star. 
These strong TiO bands may in turn be related to
the $\alpha$-element 
enhancement commonly observed in these (slightly) sub-solar metallicity stars.

\subsubsection{s-process in S stars}
After the pioneering work of \cite{Smith-1985, Smith-1986}, the abundances of the s-process elements have been investigated in S stars in \cite{Wallerstein-2003, Wallerstein-2011, Neyskens-2015, Shetye-2019, Shetye-2020, Shetye-2021}. The measured s-process abundances reproduce the nucleosynthesis predictions very well (see Fig. 5 of \cite{Shetye-2021}).
The derived ratios [s/Fe] vary between 0.2 and 1.5 dex \cite{Shetye-2021}.
However, there is a problem in the simultaneous increase of carbon and s-process elements: models predict C/O to be larger than 1 at [s/Fe]$\sim $ 0.6 dex \cite{Shetye-2021}, while many S stars (which by definition have C/O$<1$) have measured [s/Fe] larger than 0.6 dex, at odds with all TDUs models. Dust obscuration of the most evolved objects could explain why some S and C stars in advanced phases of evolution are not observed, but it can not solve the discrepancy between [s/Fe] and C/O in the observed ones. The origin of this discrepancy is unclear. Possible explanations include a higher oxygen abundance in S stars, which would keep their C/O ratio below unity; line blending effects; dynamical effects in stellar atmospheres; limitations in the stellar evolution models; 
or condensation of carbon into dust (see Sect.~\ref{sec3}).

We now examine the presence of specific heavy elements probing the AGB evolution and detected in O-rich S stars.
\begin{itemize}
    
\item {\bf Technetium}

As mentioned above, technetium has been first  detected in S-type stars in a number of studies \cite{Little-Marenin-1979,Little-Marenin-1987, Vanture-1991, Barnbaum-1993, Abia-1998,
Vaneck-1999,Lebzelter-2003, Uttenthaler-2010, Neyskens-2014, Neyskens-2015, Shetye-2018, Shetye-2019, Shetye-2020, Shetye-2021, Shetye-2025}.
However, its abundance has rarely been measured because of the difficulty in modelling the \ion{Tc}{i} line region ($\sim 4200-4300$ {\AA}) in the S star spectra.

\item{\bf The Rb diagnostic and problem}

The production of Rb is governed by the probability that $^{85}$Kr and $^{86}$Rb capture a neutron before they undergo $\beta$-decay, since these isotopes constitute key branching points on the s-process path \cite{Lambert-1995,vanRaai-2012}. This branching probability strongly depends on the local neutron density, independent of temperature. The $^{87}$Rb/$^{85}$Rb isotopic ratio would, in principle, provide a direct measure of neutron density at the nucleosynthesis site; atomic lines of different Rb isotopes cannot be separated in stellar spectra. Instead, the relative abundance of Rb to neighbouring elements, particularly Zr, is used because it is also sensitive to neutron density and therefore serves as a useful diagnostic of both the stellar mass and the dominant neutron source operating during the s-process \cite{Lambert-1995, abi01, Garcia-Hernandez-2006, Yong-2008, vanRaai-2012}. 
Observationally, [Rb/Zr]$<0$ is measured in low-mass AGB stars ($\leq 3$M$_{\odot}$), indicating that the dominant neutron source is the reaction $^{13}$C($\alpha$,n)$^{16}$O \cite{Lambert-1995, Plez-1993}.
Conversely, 
[Rb/Zr]$>0$ is observed in more massive AGB stars, in which neutrons are predominantly supplied by the reaction
$^{22}$Ne($\alpha$,n)$^{25}$Mg
\cite{Garcia-Hernandez-2006}.
Using hydrostatic models, earlier analyses inferred extremely high Rb overabundances (up to $10^3-10^5$ times solar), which posed a problem for s-process nucleosynthesis models.
Rb abundances were re-examined \cite{Perez-Mesa-2017} in massive O-rich AGB stars by extending classical hydrostatic stellar atmosphere models to consider extended atmospheres that include circumstellar envelopes and mass loss. A model of the \ion{Rb}{i} $\lambda 7800$ \AA~line in expanding dusty atmospheres revealed that the circumstellar component strongly affects the line profile. Once this is considered, the derived Rb abundances are significantly lower and fall within the range predicted by the s-process nucleosynthesis driven by the $^{22}$Ne($\alpha$,n)$^{25}$Mg neutron source in massive AGB stars ($> 4$ M$_{\odot}$) undergoing hot bottom burning (HBB). Thus, the previously claimed “extreme” Rb enhancements were largely an artefact of inadequate atmospheric modelling.
The very high abundance of Rb reported in Sakurai's object (=V4334 Sagittarii, a very late thermal pulse, post-AGB star) \cite{Asplund-1999}, though accounted for in \cite{Herwig-2011}, may similarly need to be reassessed.

\end{itemize}

\subsubsection{s-process in SC and CS stars}
Attempts to analyse the chemical composition of SC stars can be found in \cite{Dominy-1986, Smith-1990, Abia-1998}.
The studied samples have solar metallicities (in agreement with the idea that most of the Galactic AGB stars are disc objects of close-to-solar metallicities) and effective temperatures between 3000 and 3500 K. The \ion{Tc}{i} line at $\lambda 5924$~\AA~is observed, as well as the enhancement of the s-process elements by a factor of $\sim$10.
Secular changes have been observed in BH Cru and LX Cyg (increase in the pulsation period from $\sim$420 to $\sim$540 days and in the pulsation amplitude), indicating that the star’s radius expanded significantly on relatively short timescales \cite{Zijlstra-2004}.
These authors argue that changes in spectral classification can originate from temperature and equilibrium chemistry, not just changes in abundance.
In fact, when the C/O ratio is very close to 1 within a few hundredths (as observed in SC and CS stars), the opacity in the atmosphere changes very abruptly, which makes the chemical analysis extremely dependent on the adopted C/O ratio; a critical revision of the chemical pattern found in these stars with the adequate up-dated (C/O ratios) atmosphere models could be useful.

\subsection{Extrinsic stars as AGB nucleosynthesis fossils}
\label{Sect:ext}

Stars that mimic AGB stars more or less successfully have been known since the 1970s.  Barium stars were soon recognised as polluted stars \cite{McClure-1980,bus95} since they were G-K giants and therefore did not evolve sufficiently to be in the AGB phase themselves. 
Instead, their peculiarities are due to mass transfer from a former AGB companion that has since evolved into a white dwarf \cite{bus95,Jorissen-2019}.
This opened the way to the recognition that there could also be polluted stars among the cooler M stars, which have been identified as extrinsic S stars \cite{Smith-1988, Jorissen-1998}. 
Extrinsic C stars have also been discovered, but at subsolar metallicity: (i) CH stars, (ii) carbon enhanced metal-poor stars with s-element enhancements (CEMP-s), (iii) possibly CEMP-rs stars with both s- and r-element enhancements. 
The reason why such extrinsic C stars have been uncovered at subsolar metallicities is probably the observational bias against the detection of extrinsic solar-abundance giant carbon stars (spectral variability making radial velocity variations challenging to analyse, Tc lines difficult to detect, see Sect.~3).
Since pollution must in most cases, from evolutionary timescale considerations, have occurred while the to-be-extrinsic object was still on the main-sequence, dwarf extrinsic stars have been searched for, with success for dwarf barium \cite{Escorza-2019} and C stars \cite[from Gaia DR3 in][]{Roulston-25} (see also \cite{Farihi-2025}), but unsuccessfully for dwarf S stars \cite{Vaneck-17}.

Although they provide less direct, real-time snapshots of s-process nucleosynthesis than intrinsic stars, extrinsic stars can nonetheless place valuable constraints on AGB nucleosynthesis (e.g. \cite{Cseh-2022}). Indeed, as in the case of post-AGB stars, their spectra are less affected by molecular line blending, which makes the determination of stellar parameters and chemical abundances comparatively easier.

\subsubsection{Extrinsicity diagnostics}
Extrinsic stars can be distinguished from their intrinsic analogues by the following:

\begin{itemize}
    \item Radial velocity variations, since they are binaries. Among all types of extrinsic giants (CEMP-s and -rs, CH, barium, and extrinsic S stars), the orbital period ranges from $\sim 100$ days to more than a century \cite{Jorissen-1998, Lucatello-2005, Starkenburg-2014, Jorissen-2016, Hansen-2016, Jorissen-2019, Dimoff-2024}, although it is unclear how significant pollution could occur at such large orbital periods, unless the orbit has enlarged after mass transfer.

    \item Presence or absence of technetium.  $^{99}$Tc
    has a half-life of only $\tau_{1/2}=2.1 \times 10^{5}$ yr, which is much shorter than the main-sequence lifetime of low-mass stars. 
Three \ion{Tc}{i} lines are useful for O-rich stars in the optical: $\lambda$4238.19, 4262.27, and 4297.06 \AA, all of which are heavily blended. 
The dichotomy Tc/non-Tc  \cite{Little-Marenin-1979} has been shown to coincide with the dichotomy non-binary/binary among S-type stars, roughly half of which are extrinsic and the other half intrinsic \cite{Vaneck-2000}.
Bitrinsic stars have even been identified: they are extrinsic stars ascending the AGB \cite{Shetye-2020}; they are binaries and Tc-rich.

    \item Niobium overabundance: The $^{93}$Zr produced by the s-process decays into mono-isotopic $^{93}$Nb in $\tau_{1/2}=1.53 \times 10^{6}$ yr. Therefore, a (solar-scaled) normal abundance of Nb is expected in intrinsic stars (where $^{93}$Zr decay has not yet occurred), while an enhanced abundance is expected in extrinsic stars. Indeed, this is measured from the spectra \cite{Neyskens-2015, Shetye-2018, Shetye-2020, Karinkuzhi-2018}.

     \item IR excess is also discriminant. From a sophisticated decision tree using several IR colour indices that robustly trace circumstellar dust excess, \cite{Chen-2023} separate extrinsic from intrinsic stars in IR colour–colour space and also identify spectral features (\ion{Zr}{i}, \ion{Ne}{ii}, H$_\alpha$, \ion{Fe}{i} and \ion{C}{i}), which appear to be statistically efficient discriminators. The physical reason why the median equivalent width of such features is 
     generally higher in extrinsic stars than in intrinsic S-type stars remains to be investigated. 

     \item Luminosity constraints: extrinsic S stars are on average less luminous than intrinsic ones, as can be seen by comparing their luminosities \cite{Shetye-2021} or their M$_K$ magnitude: virtually no intrinsic stars are found below M$_K=-6$ \cite{Chen-2023} or log 
     L/L$_{\odot}= 3.25$ \cite{Shetye-2021}. However, this separation is blurred by the fact that S stars comprise objects of different masses ($\sim 1-3$ M$_{\odot}$) and metallicities ([Fe/H]$\sim -0.5-0.0$), and that the TDU line, which is the clear limit below which no intrinsic stars should be found, has a rather tortuous path in the HR diagram (see Fig 4 of \cite{Shetye-2021}). 
\end{itemize}

\subsubsection {The Zirconium and Niobium pair in extrinsic stars} 

Mono-isotopic Nb is produced exclusively through the $\beta$-decay of $^{93}$Zr. In extrinsic S stars, the time elapsed since the end of mass transfer from the companion is considerably longer than the half-life of $^{93}$Zr
($t_{1/2}=1.53$ Myr). Therefore, the present-day abundance ratio Zr/Nb
in an extrinsic S star reflects the ratio Zr/$^{93}$Zr at the termination of the 
s-process in the former AGB companion, assuming that the s-process contribution dominates over the initial heavy-element composition, as in the case of the most strongly enriched extrinsic S stars.
The measurement of this isotopic ratio is interesting because it can be used as a thermometer since it is directly related to the neutron capture cross sections of the Zr isotopes under the assumption of local equilibrium along the s-process path. Because these neutron-capture cross sections are temperature-dependent, the corresponding 
Zr/$^{93}$Zr ratio at the s-process site is also temperature-dependent. As a consequence, the measured
Zr/Nb ratio could in principle serve as a diagnostic (independent of stellar evolution models) of the s-process temperature. 
This has been tested in (extrinsic) S-type stars in \cite{Neyskens-2015}, leading to temperatures (less than one hundred million kelvin), which require activation of the $^{13}$C$(\alpha, $n$)^{16}$O neutron source.
This technique has also been applied to other extrinsic stars (Ba, CH, CEMP, \cite{Shetye-2021, Karinkuzhi-2018, Karinkuzhi-2021}).
However, from the analysis of an extensive sample of 180 barium stars, \cite{Vilagos-2024} conclude that in the majority of stars studied, this “thermometer” yields unrealistic temperatures, indicating that the underlying assumption of a steady-state 
s-process is not satisfied. The most likely interpretation is that the observed abundances of Zr and Nb are not solely the products of the classical equilibrium s-process nucleosynthesis. Instead, at least some of these elements may have originated from non-standard neutron capture episodes or additional processes not incorporated in current AGB models. This discrepancy highlights the need to consider more complex or time-dependent nucleosynthetic pathways
\cite{Lugaro-2023}, such as variations in neutron density regimes or contributions from non-steady s- or i-process events, when interpreting the abundances of heavy elements in these stars.

\subsection{AGB stars at subsolar metallicities}
AGB stars have been identified in globular clusters (among others 47 Tuc \cite{Worley-2012, Lebzelter-2014, Johnson-2015}, NGC 362 and NGC 6388 \cite{Worley-2010},
NGC 6752 \cite{Campbell-2017}, M13, M5, M3, and M2 \cite{Garcia-Hernandez-2015}). However, 
as mentioned before, the variability
of their photospheres, 
driven by shock waves and large-scale convective motions, produce asymmetries and Doppler shifts in spectral lines.
This makes the determination of stellar parameters difficult when using hydrostatic model atmospheres. As a result,
abundance studies beyond the light elements are challenging. Very few field low-metallicity AGB stars have been detected, except maybe CS 30322-23 \cite{Masseron-2006} and the AGB carbon star IY Hya \cite{ves01, abi02}. Specific nucleosynthesis models for metal-poor AGB stars have been computed by \cite{cri11, lug12}.


\section{AGB carbon stars}\label{sec3}
\subsection{Observational background}
Carbon stars of N-type \cite{kee93} are known as normal (i.e., intrinsic) AGB carbon stars. These stars occupy the tip in the AGB spectral sequence M$\rightarrow$MS$\rightarrow$S$\rightarrow$SC$\rightarrow$C(N), thus, as mentioned in Section 1, they are the natural result of the continuous addition of carbon into the envelope throughout the TDU episodes after each TP; they are genuine TP-AGB stars. The upper and lower mass limits for an AGB star to become a carbon star are still under discussion because these limits strongly depend on the treatment of the convection and mass loss recipes used; however, there is a broad consensus to place these limits around 1.5 M$_\odot$ and 4 M$_\odot$, respectively, the lower limit strongly depends on the metallicity.  Above $\sim 4$ M$_\odot$, the bottom of the convective envelope is so hot that almost all the dredged carbon is burnt by the CN cycle, preventing the star from becoming C rich (this is the HBB phenomenon mentioned previously). Carbon stars are very luminous objects ($<$M$_{\rm{bol}}>= -5.04\pm0.55$ mag, \cite{abi22}), so they can be easily detected and studied using high-resolution spectroscopy in stellar systems belonging to the Local Group. Furthermore, because of their high luminosity and small variability in the J band, they have recently been proposed as standard candles \citep{abg23} for the local universe. They are cool objects with T$_{\rm{eff}}\leq 3500$~K, showing a huge envelope: their radius may reach values $\sim 1$ AU, so their gravity is low (log g$\sim 0.0$, typically).

Carbon stars represent a formidable challenge from a spectroscopic perspective. Because of their cool atmospheres, they show very crowded spectra due to the presence of strong molecular absorptions. These absorptions are mainly due to carbon-bearing molecules with two or three carbon atoms, such as CN, C$_2$, or HCN. Most AGB carbon stars are variable (irregular, semi-irregular, and Mira types). As a consequence, their spectra are usually affected by large-scale movements of the photosphere (stellar pulsations, shock waves.., etc.), which induce strong line asymmetries, broadening, and Doppler shifts in the spectral lines. These phenomena 
greatly hinder the chemical analysis of these stars and, in principle, require the use of dynamical atmosphere models. Although some progress has been made in this regard (e.g. \cite{eri14,sid25}), most chemical analyses still rely on one-dimensional static atmosphere models assuming LTE, similar to the O-rich AGB star analyses. This may introduce systematic errors in the determination of the surface chemical composition and may lead to wrong conclusions on the nucleosynthetic processes that occur in their interiors. Although significant advances exist in the 3D modelling of AGB star atmospheres \citep{fre17,fre23}, their impact on the abundance determination has not yet been evaluated.

Even excluding the issue of a dynamical atmosphere, the use of standard static 1D model atmospheres typically means random abundance uncertainties in the range $\pm 0.25$ to $\pm 0.5$ dex for a given [X/Fe] ratio, depending on the specific element \citep{abi02}. These large errors are mainly due to the uncertainty in the determination of the T$_{\rm{eff}}$ and the
C/O ratio, both parameters determine the shape of the spectrum. Recently \cite{gal17} showed that variations in the C/O ratio have a large impact on the structure of 3D model atmospheres in carbon enhanced metal-poor dwarf stars, and suggested that it could also significantly affect the structure of more evolved C-rich giant stars. Gravity and metallicity play a secondary role.

At optical wavelengths, the main absorbers are the CN and C$_2$ bands (including all isotopologomes). In the near infrared spectrum, CO absorptions are also found, as well as other minority species, such as CH$_4$. Below $\sim 4500$ {\AA}, AGB carbon stars show almost no flux, probably due to the absorption of C$_3$ bands.  Atomic lines are usually difficult to detect (except huge ones such as the Balmer H lines, the Na-K lines, and some K lines), even if using high-resolution spectroscopy ($R>40,000$). Only a few spectral windows are suitable for abundance analyses, provided a complete list
of C-bearing molecules is included in any spectral line list. These windows are located in the spectral ranges $4700-5000$ {\AA}, $\sim 7800$ {\AA} and $8000-8200$ {\AA}. In the near infrared spectrum (0.9-2.5 $\mu$m), some atomic lines are useful for abundance analysis. However, much effort has still to be put into identifying atomic spectral lines of
promising utility in this latter spectral region (see, e.g. \cite{zen23}). 

\begin{figure}[ht]
\centering
\includegraphics[width=0.99\textwidth]{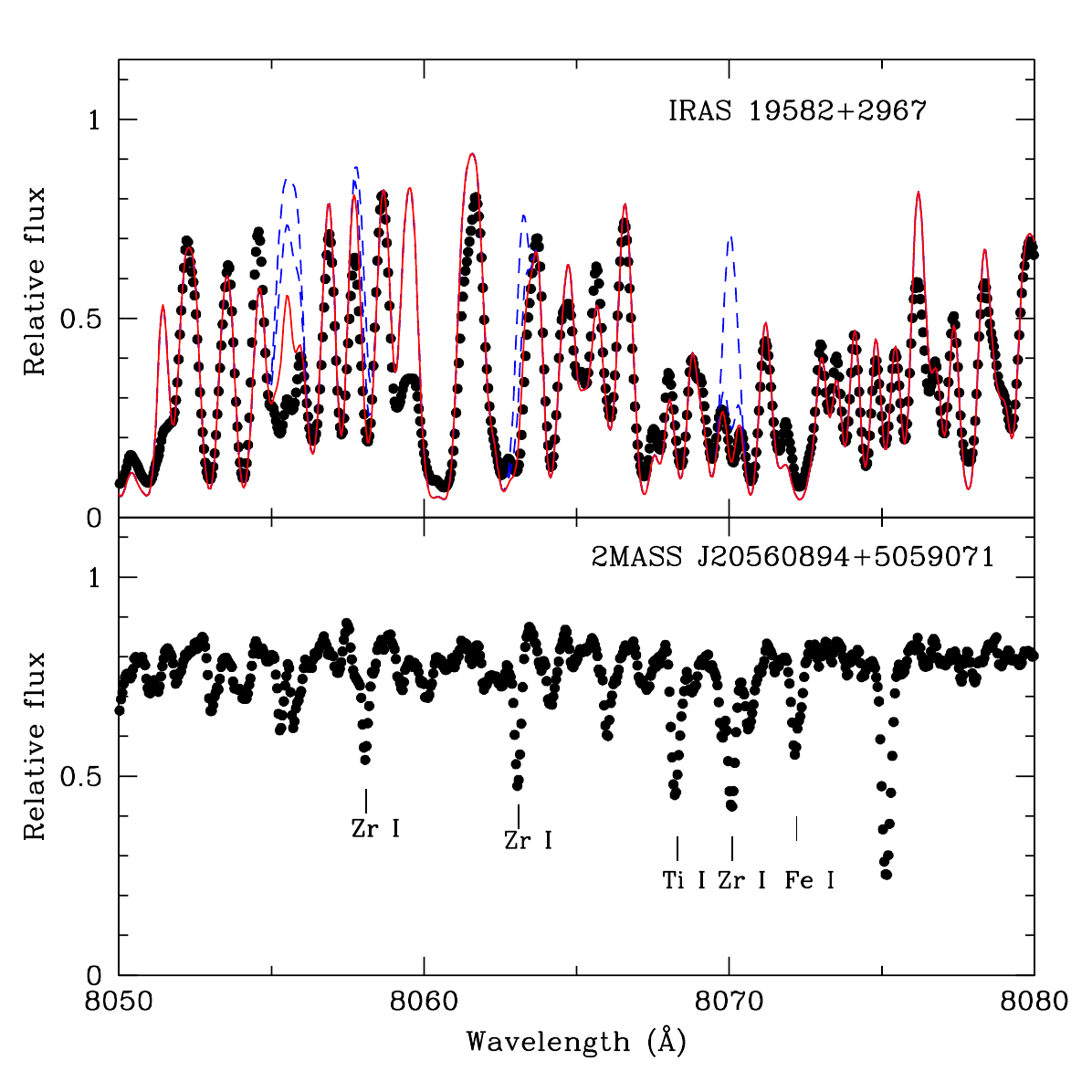}
\caption{Comparison of observed (solid black circles) and synthetic spectra (red and blue lines) of the two TP AGB stars. The bottom panel shows the O-rich AGB star 2MASS J20560894+5059071, where some atomic lines are marked. 
Both stars have near-solar metallicity.
The top panel shows the same spectral region except for the carbon star IRAS 19582+2967. Note the difference in the observed spectrum in this spectral region between an O-rich and a C-rich star; the former star is dominated by TiO and atomic absorptions, whereas the C star is dominated by CN lines (in fact almost all the absorptions seen in the top figure are due to CN). The dashed blue lines in the carbon star are the synthetic spectra with no-Zr and [Zr/Fe]=0.0 dex, and [Zr/Fe]=+0.6 dex (continuous red line).  Figure adapted with permission from reference \cite{abi25}; copyright 2025 A\&A.}\label{fig2}
\end{figure}

Figure 4 shows the observed spectra in the $\sim 8000$ {\AA} region in a typical AGB carbon star (top panel) and an O-rich AGB star (bottom panel) as an illustrative example of the difference between the spectra of a C-rich star and an O-rich AGB star. Note that both stars have very similar effective temperatures, gravity, and metallicity; they only differ in the C/O ratio. Some atomic lines detected in the O-rich star are marked and appear to be heavily blended in the carbon star by CN absorptions. However, they can still be used in carbon stars, providing 
high-resolution spectra, and spectral synthesis is used. Clearly, the actual C/O ratio determines the shape of the spectrum: the higher the ratio, the larger the absorption of the carbon-bearing molecules. Interestingly, the C/O ratio determined in the overwhelming majority of Galactic carbon stars (see, e.g. \cite{lam86,abi02}) typically does not exceed largely unity (C/O $\leq 1.6$). This is inconsistent with standard theoretical expectations (see, e.g \cite{kar14, cri11}), which predict much larger C/O and $^{12}$C/$^{13}$C ratios in the envelope than those observed in an AGB star when it becomes C-rich. The reason for this discrepancy is still unknown, which shows our limited knowledge of the modelling of the mixing processes occurring in AGB stars, particularly the TDU and mass loss phenomena (e.g \cite{ven16}). Nevertheless, some scenarios have been proposed to explain this discrepancy. For example, \cite{lod17} suggested that as carbon is transported to the surface by the action of the TDU episodes and the star progressively cools in its ascent to the AGB phase, the carbon atoms condense into amorphous graphite, TiC, and SiC grains on the cool surface of the star. Consequently, the gas phase at the surface is continuously depleted in carbon by condensation into grains. This may cause the C/O ratio in the gas phase (what we actually see in the visual and near-infrared spectra) to never exceed unity significantly. This process is highly dependent on the star's metallicity: in metal-poor carbon stars, C/O ratios largely exceeding unity are found, as is the case in the few extragalactic metal-poor AGB carbon stars studied to date \citep{del06, abi08}. In any case, even for metal-poor carbon stars, the observed C/O ($^{12}$C/$^{13}$C) ratio in AGB carbon stars disagrees with current theoretical predictions.  It is worth stating here that stellar grains of
type Y (representing a few percent of the observed SiC grains), which are thought to form in metal-poor C-rich AGB stars, typically show $^{12}$C/$^{13}$C$>100$, (in some cases even $>1000$ \cite{nan25}); these high carbon isotopic ratios would agree better with theoretical predictions, but it further highlights the current discrepancy with the $^{12}$C/$^{13}$C observed in the atmospheres of the majority of these stars. 

N-type stars are intrinsic, but the existence of N-type stars that were previously extrinsic O-rich objects cannot be excluded. Simple evolutionary reasoning shows that the formation of an extrinsic AGB carbon star is difficult at solar metallicity \citep{abi02}. Indirect evidence that N-type stars are in the TP AGB phase can be obtained from the Gaia-2MASS diagram (see above). In this diagram, the TP AGB carbon stars are located in a very specific area. The C-rich nature of many Galactic AGB stars has been found in this way and was later confirmed spectroscopically (\cite{leb23, abi22, abi25}). Obviously, an unambiguous answer about their intrinsic or extrinsic nature can only be obtained from the detection of the radioactive $s$-nuclide $^{99}$Tc, a signature of the in situ operation of the s-process in the star. This is not an easy task in N-type stars because the strongest Tc lines are placed in a very crowded spectral region ($\lambda\sim  4260$
{\AA}), where  N-type stars are  quite  faint.  
In carbon stars, the only available Tc line is that at $\lambda\sim  5924$  {\AA}. \cite{abi01,abi02} found in a sample of Galactic N-type stars that $\sim 60\%$ of them clearly showed this line; however, due to the difficulty in analysing this spectral region, the presence of Tc in the rest of the sample could not be excluded.

\subsection{s-process in carbon stars}
AGB stars are believed to be the main source of s-elements in the universe, at least those belonging to the main s component. The first detailed chemical analysis of s-elements in AGB carbon stars by K. Utsumi \cite{uts70, uts85} was based on low-resolution spectra. Utsumi found that N-type stars in the Galaxy were typically of solar  metallicity, presenting mean $s$-process
element enhancements of a factor $\sim 10$. This figure was accepted as an extreme result of the $s$-element enhancement during the AGB phase.
However, later and more accurate studies based on high-resolution and high signal-to-noise spectra 
in a more extended sample of N-type stars \citep{abi01,abi02,abi08,abi19,abi25,del06}, let to strong revision in the quantitative $s$-element abundances. N-type stars were confirmed to have near solar metallicity, but they show on average $<$[ls/Fe]$>=+0.67\pm 0.10$ dex and $<$[hs/Fe]$>=+0.52\pm 0.29$ dex,
which is approximately a factor of two lower than that estimated by Utsumi. These enhancement values are of the same order as those derived from less-evolved intrinsic S-type stars with similar metallicity (see Section 2). Contrary to the superficial appearance, this does not contradict the expected s-process enhancement along the AGB phase. In fact,
one has to consider that the sample of N-type stars studied in the above mentioned works typically shows C/O ratios that are slightly larger than unity. Theoretically, S-type AGB stars ($0.5<$C/O$<1$) can become a C-rich object in very few (one or two) TP and TDU episodes. This is sufficient to drastically change the appearance of the star because above C/O $=1$, bands of carbon-based molecules become dominant (as discussed in Section 1), but the enhancement of the $s$-process elements in the envelope is changed only very slightly. This agrees with the observations. 
Later, as the star evolves in the AGB phase and, TPs and, TDUs continue to operate, more carbon and $s$-elements are dredged into the envelope.
Theoretically, s-element enhancements by a factor of ten with respect to the Sun are found
at the very last TPs in standard AGB nucleosynthesis models of $\sim 1.5-4.0$ M$_\odot$ stars and solar 
metallicity (e.g., \cite{cri11,cris15,kar16}). Observationally, 
there is a gap in the level of $s$-element and carbon enhancement between N-type stars and their descendants: the post-AGB stars, in which s-element enhancements larger than a factor of ten are usually found \citep{van00,dem15,dem16}). In addition, as previously discussed, the increase in the mass loss rate at the very end of the AGB phase may form a thick and dusty circumstellar envelope that obscures the star at optical wavelengths; meanwhile, TPs and TDUs continue to work, continuously increasing the amount of carbon and s-element dredged up to the envelope. We do not observe these dusty and obscure carbon stars. However, post-AGB stars may undergo nucleosynthesis beyond the end of the AGB phase; thus, their abundances may not reflect the predicted surface abundances at the end of the AGB phase.

\begin{figure}[ht]
\centering
\includegraphics[width=0.99\textwidth]{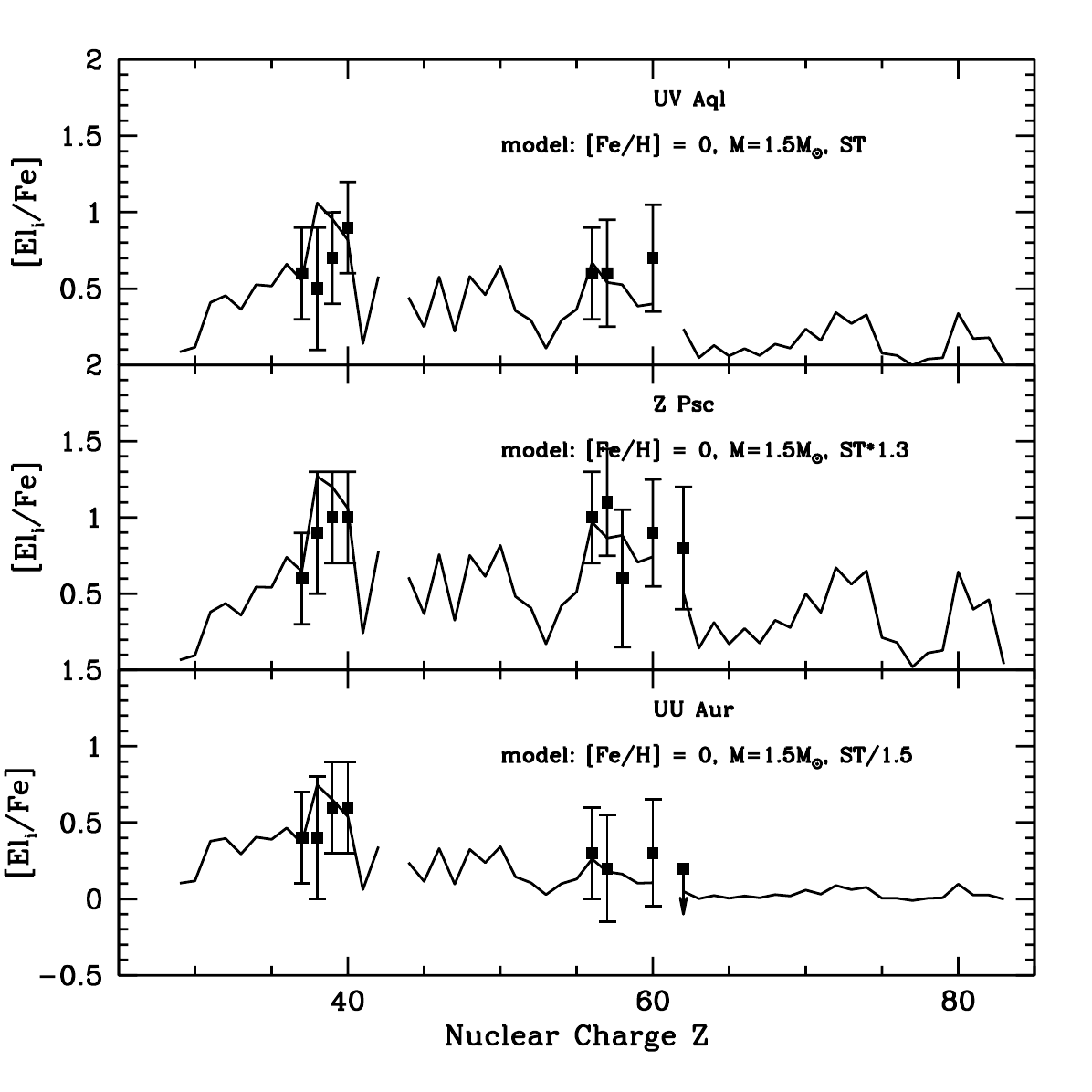}
\caption{Detailed reproduction of  
observed abundance ratios of several 
s-elements (noted as [El$_i$/Fe]) in the Galactic carbon stars UV Aql, Z Psc, and UU Aur. Theoretical models (solid lines) are computed with the FUNS code for a 1.5 M$_\odot$, solar metallicity AGB model using different choices for the size of the $^{13}$C pocket with respect to the standard (ST) choice (see \citep{abi02} for details on the properties of this ST $^{13}$C pocket). Figure adapted with permission from reference \cite{abi02}; copyright 2002 ApJ. }\label{fig4}
\end{figure}

The interpretation of the N-type star abundances in the framework of the low-mass star evolution along the AGB is confirmed and strengthened by an analysis of the
detailed element distributions in the observed stars. Figure 5 shows a sample of fits to the abundances determined in individual Galactic carbon stars, obtained by comparing their observed composition with that yielded (continuous lines) by the theoretical models in the photosphere at C/O$= 1$. Theoretical predictions correspond to 1.5 M$_\odot$, Z$=Z_\odot$ stellar models with different choices for the size of the $^{13}$C pocket formed (see details in \cite{gal88,abi02}). The plots show satisfactory agreement, confirming that most, if not all, N-type stars of the Galactic disc are indeed of low initial mass. In fact, using models of intermediate mass stars (M$> 4$ M$_\odot$), would often be incapable of reproducing at least the ls zone, resulting in a high [Rb/Sr] ratio \citep{abi01,gar09,vanRaai-2012,per17}. Similar plots for the few extragalactic metal-poor AGB carbon stars analysed to date also show a comfortable agreement with theoretical predictions \citep{abi08p}. This conclusion for the mass of AGB carbon stars agrees with the present independent understanding of the different classes of carbon stars, provided by astrometric, kinematic, and photometric criteria \citep{ber02,abi20,abi22,leb23,Roulston-25}.  Another indirect determination of the masses of AGB C-rich stars would be the analysis of possible binary systems, but this is a very difficult task since stellar pulsation, large scale convective surface movements, and the crowdedness of their spectra make it very difficult to measure precise radial velocity variations associated with binarity.

A relevant parameter in the s-process is the neutron exposure, i.e. the product of the neutron density and thermal velocity of the neutrons integrated within the period in which the seed nuclei are exposed to neutrons. An estimate of the average value of this parameter can be determined observationally through the [hs/ls] ratio, which is the relative production of the second to the first s-process peak, and it is expected to increase with decreasing stellar metallicity. This is because the total amount of neutron released by the $^{13}$C$(\alpha,n)^{16}$O source depends on the amount of $^{13}$C in the pocket located on top of the He-rich mantel, which is  marginally dependent on the initial metallicity. In fact, the amount of $^{13}$C left by the H-burning shell is not sufficient to drive a suitable s-process nucleosynthesis, so that the activation of the main neutron source requires a primary production of $^{13}$C, as due to the triple-alpha conversion into $^{12}$C, eventually followed by the $^{12}$C$(p,\gamma)^{13}$N$(\beta^{+}\nu)^{13}$C. On the contrary, the metallicity determines the amount of iron in the $^{13}$C pocket, which is the main s-process seed. In practice, the lower the metallicity, the higher the neutron-to-seed ratio and, in turn, the larger the production of heavy s-elements. The overall result is that the number of neutrons captured by each iron seed is larger and that the s-process can reach the second peak more easily. Below 
[Fe/H]$\sim -1.0$, the third peak is also reached and the [hs/ls] ratio becomes roughly constant because the first and second peaks are in steady-state equilibrium.

Figure 6 shows the comparison of the [hs/ls] ratio versus the metallicity of AGB carbon stars, where this ratio has been determined \citep{abi02,abi08,abi25,del06} with  theoretical predictions of s-process nucleosynthesis in AGB stars according to the FUNS code for different stellar masses \citep{cri11,cris15}. The continuous line represents the prediction for a  1.5 M$_\odot$ model; short dashed line represents the 2 M$_\odot$ model, while the long dashed line represents the 3 M$_\odot$ model. Note that in this case, a different standard $^{13}$C pocket \citep{cri11,cris15} was adopted compared to that used in Figure 5. The theoretical [hs/ls] ratios  shown are those when C/O$\sim 1$ is reached in the envelope. The theory and observations agree well, despite the large observational uncertainty. Nucleosynthesis calculations using different stellar codes \citep{lug03,lug12,sie13,kar14}, give very similar results for the same stellar mass and metallicity models. Nevertheless, for each [Fe/H], there is clearly a spread in the models and data, even considering the observational errors. This theoretical spread is also found when other prescriptions are used for the formation of the $^{13}$C pocket. For example, the NUGrid collaboration \citep{her07,bat19} predicts ratios higher than those of other models due to overshoot at the TP base. On the other hand, theoretical models that include the formation of the $^{13}$C pocket via magnetic buoyancy \citep{bus21} predict ratios lower than the other models because this mechanism results in a flatter and lower abundance profile of $^{13}$C. The models by \citep{kar16} also show relatively large variations depending on the initial mass due to the effects of temperature on the activation of the $^{13}$C and the $^{22}$Ne neutron sources. Other effects not shown in Fig. 4, such as rotation and diffusive mixing, may also play a role, and it is difficult to disentangle, which may play a major role in the observed spread \cite{her03,sie04,pie13}. 

Recently, the same figure has been found in extrinsic Ba-stars \citep{ror21,ror23}. Although the process of binary mass transfer is still not fully understood, the ratios of two elemental abundances are affected by the same dilution during the mass transfer, so the observed [hs/ls] ratios can be safely compared with the theoretical predictions \citep{dim25}. Because the observational uncertainties in the analysis of Ba stars are considerably lower than those in AGB carbon stars, one may conclude that this spread in the [hs/ls] exists in nature simply because the amount of $^{13}$C burnt may vary depending on the stellar mass and metallicity or other still unknown physical parameters. This observation confirms that the second peak s-elements are preferentially produced with decreasing metallicity with respect to the first peak, and the $^{13}$C$(\alpha,n)^{16}$O is the main source of neutrons in carbon stars. Figure 6 also shows that the observed [hs/ls] seems to saturate at [Fe/H]$\sim -1.0$, which also agrees with the theoretical predictions. 
Unfortunately, the available Pb spectral lines (a genuine element of the third s-peak) cannot be detected in AGB carbon stars, which prevents inferring the [Pb/hs] ratio with metallicity in these stars. Nevertheless, in extrinsic carbon enhanced metal-poor stars, the high [Pb/Fe] and [Pb/hs] ratios measured agree with the theoretical expectation that for decreasing metallicity, the s-process populates the higher mass nuclei in the stability valley, independent of the recipe used for the formation of the $^{13}$C pocket \citep{vane03,ham19,cou24,kar16,bus21}. Nevertheless, abundance studies of s-process enriched post-Asymptotic Giant Branch (post-AGB) stars \citep{kam21} in the Galaxy and the Magellanic Clouds show a discrepancy between observed and predicted Pb abundances as they result much lower than predicted. Some theoretical studies have pointed to the occurrence of the i-process (a neutron capture nucleosynthesis process with neutron density intermediate between s- and r-process \citep{van24}) to explain the observed Pb patterns in these post-AGB stars. However, a recent study by \cite{men25} in the post-AGB star 
J003643.94--723722.1 in the SMC, seems to reconcile Pb observations with standard theoretical predictions of the s-process.

\begin{figure}[ht]
\centering
\includegraphics[width=0.99\textwidth]{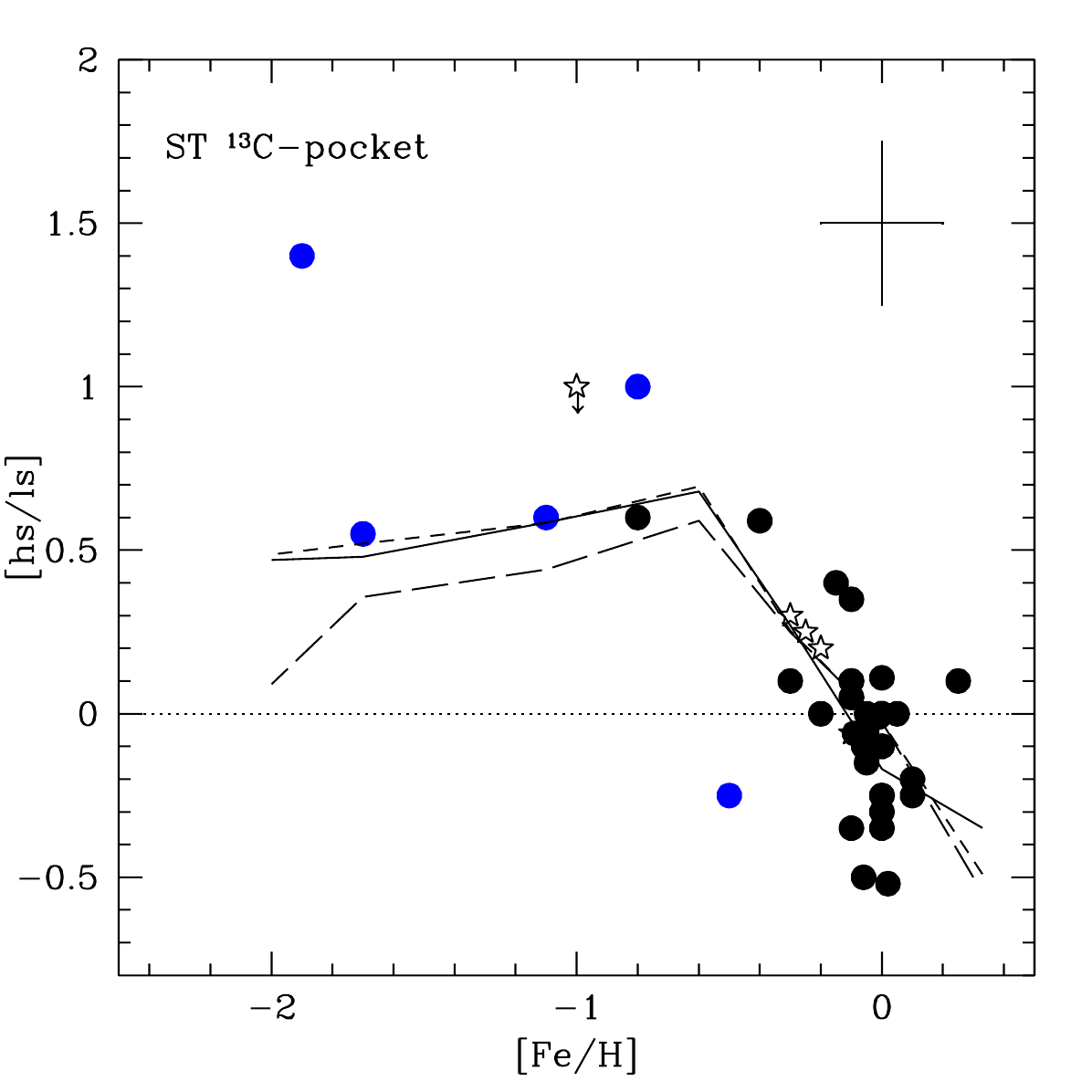}
\caption{Comparison of observed mean
low-mass (Y and Zr) with respect to the high-mass (Ba, La, Ce, and Nd) s-elements ([hs/ls]) versus metallicity for the AGB carbon stars chemically analysed up to date. The black solid dots are Galactic N-type stars; the blue solid dots are extragalactic carbon stars, and the open stars are N-type stars belonging to Galactic open clusters (see \cite{abi02,abi08,abi25,del06}). Observations are compared with theoretical predictions for a 1.5, 2.0 and 3.0 M$_\odot$ TP-AGB models (continuous, dashed, and long dashed lines, respectively) assuming a standard (ST) $^{13}$C pocket choice following the AGB nucleosynthesis models by \cite{cri11, cris15}}. The theoretical predictions are for the stellar models at the last TP. Several stars coincide in the  same data point. A typical error bar is shown. {\label{fig6}}
\end{figure}

A similar comparison with the theoretical predictions for intermediate-mass AGB stars (4-6 M$_\odot$) does not show this broad agreement with the observations. In the rarer AGB stars of intermediate mass, $^{22}$Ne would be instead preferred as a neutron donor because of the high ($\sim 3\times 10{^8}$ K) temperature in TPs. Furthermore, the $^{13}$C pocket and the hydrodynamic penetration of protons in the upper layers of the He inter-shell are less favoured. Therefore, the standard case for the choice of the $^{13}$C pocket in the intermediate-mass case is less efficient by an order of magnitude than in low-mass AGBs. Different compositions are expected due to the very different neutron densities provided by the two neutron-producing
reactions, especially for the ratio between Rb (whose isotope $^{87}$Rb is
strongly fed only at high neutron densities) and its neighbours Sr, Y, and Zr. On this basis \cite{abi01} showed
that N-type stars are typically of low mass (M$\leq 3$ M$_\odot$).

\section{Summary and future work}\label{}
During the last few decades, employing advanced spectroscopic techniques and high-precision measurements, the abundance of s-elements has been determined in a plethora of Galactic O-rich and C-rich AGB stars. These abundance determinations confirm the nucleosynthesis diagnostics obtained from independent stellar models and nucleosynthesis calculations. Future large surveys hold significant potential for uncovering a wealth of new intrinsic (and extrinsic) AGB stars in our Galaxy and in the stellar systems of the Local Group, yet to be analysed. This would allow us to really understand, among other things, the extreme dependence of the s-process on stellar metallicity. This underlines the relevance of both large spectroscopic surveys with a large number of objects and detailed studies of key targets to progress in our understanding of the s-process
nucleosynthesis. However, a desirable tool for doing this would be the accurate identification and characterisation of new atomic and molecular lines in the near infrared spectrum, where the spectra of AGB stars appear more accessible for abundance analyses. Furthermore, the use of 3D dynamical atmosphere models would be desirable in future abundance analyses of AGB stars.

\backmatter





\bmhead{Acknowledgements}
This work was partially supported by the Spanish project PID2021-123110NB-I00 financed by MICIU/AEI/10.13039/501100011033, by FEDER, una manera de hacer Europa, UE and by Fondation ULB. CA dedicates this work to her life partner Inma Dom\'\i nguez, for her unconditional support, loyalty, and love throughout her life. CA is deeply grateful for the opportunity to contribute to this volume in memory of Roberto Gallino, a friend and mentor.
SVE is likewise honoured to join in this tribute to a scientist whose work has profoundly shaped the field.

\section*{Declarations}
Not applicable

\bibliography{sn-bibliography}

\clearpage
\newpage

\begin{figure}[ht]
\centering
\includegraphics[width=0.99\textwidth]{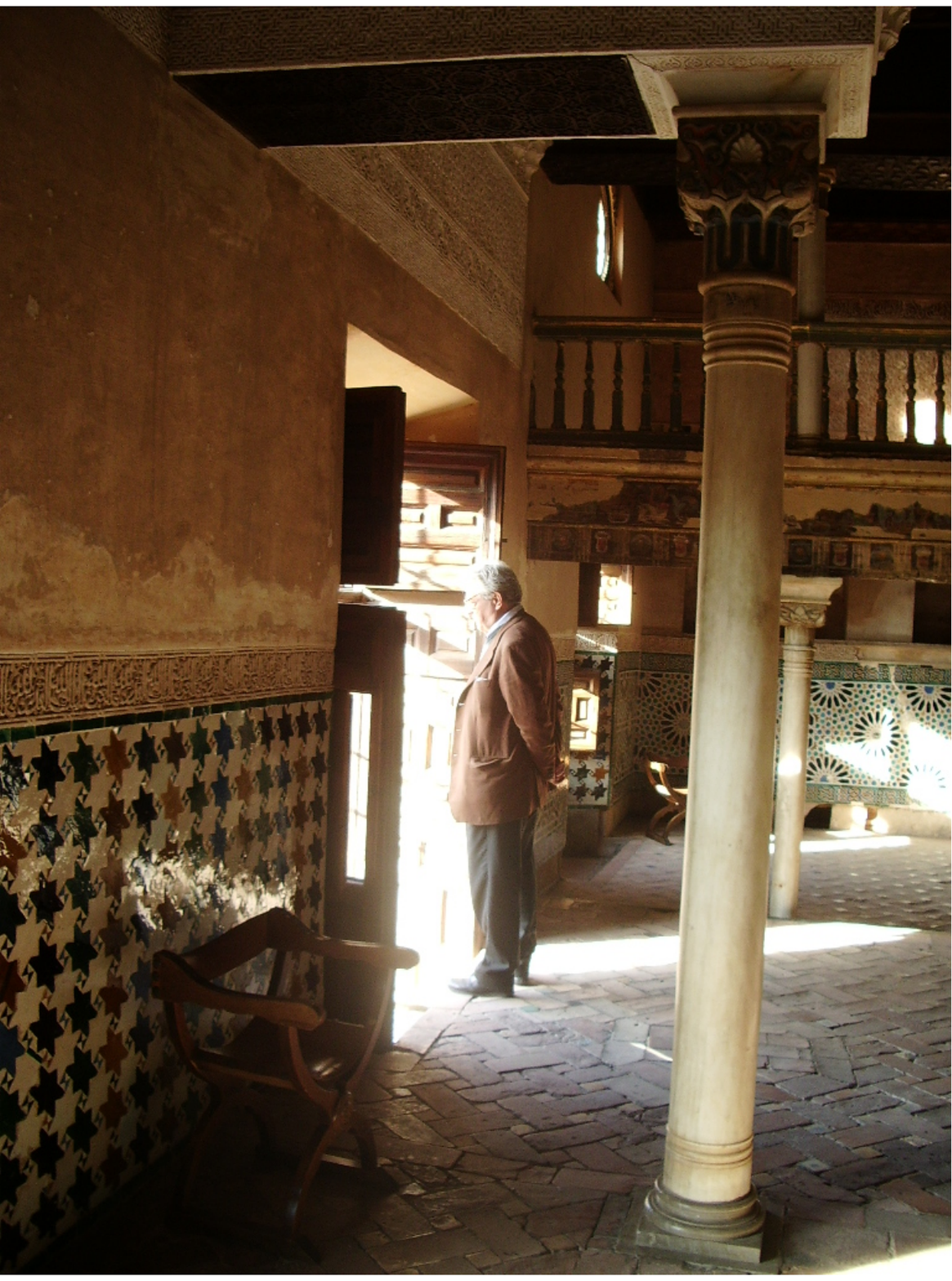}
\caption{Prof. Roberto Gallino during the visit to the Alhambra Palace at the VIII Torino Workshop on nucleosynthesis in AGB stars held in 2006 in Granada (Spain).}\label{Roberto}
\end{figure}
\end{document}